\documentclass[aps, draft, floats,showpacs,nofootinbib]{revtex4}
\usepackage{epsf}

\advance \voffset .5 in

\def\tlgo{tree-level generated operator}
\def\qwe{f}
\def\mexp{M_{\rm exp}}
\def\sm{Standard Model}
\def\gesim{\,{\raise-3pt\hbox{$\sim$}}\!\!\!\!\!{\raise2pt\hbox{$>$}}\,}
\def\lesim{\,{\raise-3pt\hbox{$\sim$}}\!\!\!\!\!{\raise2pt\hbox{$<$}}\,}
\def\tev{\hbox{TeV}}
\def\gev{\hbox{GeV}}
\def\ie{{\it i.e.}}
\def\su#1{{SU(#1)}}
\def\ui{U(1)}
\newcommand{\nc}{\newcommand}
\nc{\beq}{\begin{equation}}  \nc{\eeq}{\end{equation}}
\nc{\bea}{\begin{eqnarray}}  \nc{\eea}{\end{eqnarray}}
\nc{\baa}{\begin{array}}     \nc{\eaa}{\end{array}}
\nc{\bit}{\begin{itemize}}   \nc{\eit}{\end{itemize}}
\nc{\ben}{\begin{enumerate}} \nc{\een}{\end{enumerate}}
\nc{\bce}{\begin{center}}    \nc{\ece}{\end{center}}
\def\ssb{spontaneous symmetry breaking}
\def\vev{vacuum expectation value}
\def\ocal{{\cal O}}
\def\up#1{^{\left( #1 \right) }}
\def\rcal{{\cal R}}
\def\gcal{{\cal G}}
\def\hcal{{\cal H}}

\begin{document}

\begin{flushright}
UCRHEP-T360\\
July, 2003
\end{flushright}

\title{Natural and model-independent
conditions for evading the limits on the scale of new physics}
\author{ Jos\'{e} Wudka}\email{jose.wudka@ucr.edu}
\affiliation{Department of Physics \\
University of California-Riverside \\
California, 92521-0413\\
\vspace{.3cm}
}
\date{\today}


\begin{abstract}
One of the most interesting phenomenological quantities connected 
the physics expected to underlie the \sm\ is its scale $\Lambda $. In this paper 
I argue that the limits on this quantity obtained using 
model-independent parameterizations contain an tacit assumption that
could be invalidated under a variety of situations. As a specific example,
existing limits on $\Lambda $ would be decreased by at least an order 
of magnitude if the underlying physics has a symmetry under which all \sm\
particles are singlets but none of the heavy excitations are. In this
case current experiments would see no clear indications of new physics
using precision measurements while future colliders capable of
directly producing the heavy particles would only occur in pairs.
\end{abstract}

\maketitle

\section{Introduction}
One of the central issues in the many attempts to describe the physics underlying
the \sm\  is the determination of the
scale at which these prophesied interactions will be directly
observed. To date no (unambiguous) non-\sm\ effects have been found either through
direct production, or indirectly through virtual effects\cite{limits}; 
accordingly the data
has been used to provide limits on the scales and couplings of these 
interactions\cite{limits.models,limits.leff}.
As this data becomes more precise and experiments probe higher energies, the limits on
the scale of new interactions is constantly being pushed to higher values. In some
cases these limits have become so stringent that the scales of new physics
are stated to lie beyond the reach (through direct observation) of near-future 
accelerators such as the LHC~\footnote{For example, the limits on 
the scale of the interactions
responsible for 4-fermion operators involving two quarks and 
two leptons (all left-handed) is $ \gesim25\tev$~\cite{pdg}}. 

In many instances limits on the scale(s) associated with the new interactions are
obtained within the context of specific models~\cite{limits.models}, but an alternative
non-model specific approach has also been repeatedly
followed~\cite{limits.leff}, where these limits
are obtained using generic couplings limited only by naturality 
constraints~\cite{patterns}.

Within this ``generic'' approach it is possible to understand the absence of
non-\sm\ effects by the assumption that the scale of the new interactions
is so large it lies outside the sensitivity limit
of all current experiments (involving real or virtual heavy 
excitations)~\footnote{This assumption is consistent whenever 
the heavy physics decouples~\cite{decoupling};
if it does not the scale of the new physics will be directly accessible at the 
LHC~\cite{non.dec}.}. But
an alternative possibility, which will be examined below,
is that {\it(i)} the masses of the heavy excitations are too large
to have been directly produced at the energies available, and
{\it(ii)} that the couplings in the underlying
theory are naturally suppressed so as to insure that current experiments
are blind to all leading virtual processes involving the heavy physics. 
Under these conditions current 
precision data would  exhibit only very small deviations from the \sm\
predictions, even when the scale of new physics is quite close to the energies currently
being probed directly, and near-future colliders would produce new particles without any 
premonition of their existence form LEP or Tevatron data.

In this paper I will investigate the conditions under which this last possibility can be realized.
For definiteness I will consider the case where the physics underlying the \sm\ 
is decoupling and weakly coupled (for
non-decoupling theories, the scale of new physics is constrained by the consistency of the
theory to be of  order  $ 4\pi v \sim 3$TeV, where $v$ denotes the \sm\
vacuum expectation value~\cite{non.dec}). To this end I will first
parameterize the new physics effects by an effective
Lagrangian~\cite{leff}, then I will characterize the leading contributions
(Sect. \ref{sect:leff}), and use this to provide constraints
that would naturally insure their absence (Sect. \ref{sect:eliminate}). 
Parting comments and some numerical estimates are presented in section
\ref{sect:end}. A calculational detail is relegated to the appendix.

\section{Effective theory and \tlgo s}
\label{sect:leff}

I will assume the existence of non-\sm\ interactions that can be described by
a gauge theory
whose scale $\Lambda$ lies significantly above the energies currently probed
directly. Furthermore I will restrict the discussion to the case where the
corresponding new interactions are weakly coupled, and such that they decouple
in the limit $ \Lambda \to \infty $~\cite{decoupling}. In this context the \sm\ 
is the low-energy limit of this more fundamental  theory; I will
refer to all \sm\ particles as ``light'', and assume that all the
other excitations have a mass $ \gesim \Lambda $.

At low-energies (\ie\ below $ \Lambda$) the heavy excitations 
will manifest themselves through
virtual processes. These effects can be parameterized by a
set of effective operators involving only light fields
and respecting the local symmetry of the \sm.~\cite{leff}
The condition that
the heavy physics decouples implies that all observable effects (that is, all
the virtual effects that cannot be absorbed in a renormalization of the \sm\
parameters) will disappear in the
limit where the scale of the heavy physics $ \Lambda $ is 
sent to infinity.~\cite{decoupling}

If the heavy
physics is also weakly coupled, then the larger the (canonical) dimension of an
operator the smaller its impact. This is so because the anomalous
dimensions of all operators will be small, so that to a good 
approximation an operator
of canonical dimension $n$ will appear with a coefficient proportional to 
$ \Lambda^{4-n}$, and its contribution to a process with characteristic energy $E$ 
will be suppressed by a factor $ \sim (E/\Lambda)^{4-n}$~\cite{leff}. Note that 
$E \ll \Lambda $ is a tacit but central 
(and unavoidable) assumption when using an effective Lagrangian parameterization.

This observation generates a simple hierarchy among effective operators. The
leading effects are generated by operators of dimension $ \le4$ that respect the 
$ \ui_Y \times \su2_L \times \su3_c $ gauge symmetry; by definition these
correspond to the terms in \sm\  Lagrangian. The most important
subleading terms are generated by
operators of dimension 5 and 6, which have been listed.\cite{weinberg,bw}

Naturality~\cite{nat} constraints can now be used to further refine the above 
hierarchy, which is important when considering the 
phenomenological applications of this approach.
An operator of dimension $n$ appears with a coefficient of the form 
$ \qwe/\Lambda^{n-4} $ where $\qwe$ is a numerical factor. When
the underlying theory is weakly coupled, $\qwe$ will equal a 
product of coupling constants for those operators that can be generated at tree
level.~\footnote{Coefficients generated at tree level will also receive
radiative corrections, but these are small by assumption.}
In contrast, the coefficients for operators that are only 
generated via loops contain
additional coupling constants
and a numerical suppression factor $ \sim 1/(4\pi)^2 $. 

It  follows that in the absence of  \tlgo s 
the leading virtual effects produced by the heavy
physics (to processes allowed within the \sm)
will be of order $ 1/[ G_F ( 4 \pi \Lambda)^2 ] $, which is
of the same order as the \sm\ radiative effects {\em supressed}
by an additional factor of $(m_W/\Lambda)^2 $. In this case the limits on
$ \Lambda$ obtained form precision data will not improve on
the direct ones.

The diagrams responsible for \tlgo s can be characterized as 
follows.\footnote{In extracting the conditions for the absence of \tlgo s I will ignore
any operator of dimension $ \ge5$ in the {\em underlying} theory. This
is reasonable not because these terms are necessarily absent, but because,
if present, they would be the result of the virtual effects of some yet
heavier physics with scale $ \Lambda' \gg \Lambda $. The
coefficients of these operators will then 
be then suppressed by powers of $ \Lambda/\Lambda' \ll 1$.}
Consider an arbitrary graph with
$I$ internal lines, $E$ external lines and vertices labeled $V_n$, and let
$h_n$ be the number of heavy legs in $V_n$. For the situations considered,
all internal lines are heavy and all external lines are light. If the graph
contains $L$ loops then we have the relations
\beq 
L = I + 1 - \sum V_n; \quad \sum h_n V_n = 2 I \quad \Rightarrow \quad
\sum(h_n-2) V_n = 2 (L-1)
\eeq 
Tree level graphs ($L=0$) must then contain at least one vertex with 
$h_n =1$.

A theory will produce no \tlgo s when the heavy excitations are integrated
out if and only if the Lagrangian (after \ssb) contains no terms with a single 
heavy field, that is, if the following vertices are absent:
\beq
\psi \psi \Phi
\quad \phi \phi \Phi
\quad \phi \phi \phi \Phi
\quad \psi \phi \Psi
\quad \psi \psi X 
\quad \psi A \Psi
\quad \phi \phi X
\label{eq:tlv}
\eeq
($ \Psi,~\Phi,~X$ denote heavy fermions, scalars and vector bosons respectively, and
$\psi,~\phi,~A$ their light counterparts); the vertex $XA \phi \phi $, 
is not included since it follows form the absence of vertex $ X \phi \phi $ (see the appendix).

The list (\ref{eq:tlv}) contains only vertices of dimension $\le4$. This is not because 
higher-dimensional terms are necessarily absent in the theory underlying the \sm, but
because, if present, they would result form virtual interactions whose typical scale $ \Lambda'$
is much larger than $\Lambda$. The coefficients of these higher-dimensional terms would be
the suppressed by powers of $ \Lambda/\Lambda' \ll 1 $ and can be ignored.

The elimination of \tlgo s by excluding the list in (\ref{eq:tlv})
can of course be restricted to
vertices satisfying certain symmetry properties, such as, for example,
being baryon or lepton number conserving.
Note also that requiring the
absence of \tlgo s of dimension 5 and 6 implies the absence of \tlgo s 
to all orders in $1/\Lambda $.

\subsection{Operators of dimension 5 and 6}
\label{sect:dimvvi}

It is straightforward to vertify these general arguments for the case of dimension
5 and 6 operators. Assuming the particle content of the \sm\ with a single scalar
doublet, and with the addition of 
right-handed neutrinos, all gauge-invariant operators of dimension 
5 are necessarily lepton-number violating~\cite{weinberg}:
\beq
\ocal\up5_1=
\left( \bar\ell \tilde \phi \right) \left(\phi^\dagger \ell^c \right)
\qquad
\ocal\up5_2=
\left( \bar\nu \nu^c \right) \left( \phi^\dagger \phi \right)
\eeq
where $ \ell $ denotes a left-handed fermion iso-doublet, $ \phi $ the \sm\
scalar doublet (and $ \tilde \phi = i \sigma^2 \phi^*$), $ \nu$ a  
right-handed neutrino singlet, and $ \sigma^I$ the
usual Pauli matrices; family indices have been suppressed and the superscript
$c$ denotes the charge-conjugate fields.

$ \ocal\up5_1 $ can be generated at tree level  by the exchange of a heavy
scalar $\su2_L$ triplet of unit hypercharge~\footnote{The conventions used 
are such that $ \phi,~\ell$ and $e$ have hypercharges $1/2,~-1/2$ and $-1$ 
respectively} or by a zero-hypercharge heavy fermion singlet or triplet. $ 
\ocal\up5_2$ can  be generated by the exchange of a heavy scalar singlet 
of zero hypercharge, or of a fermion triplet of hypercharge $1/2$. The 
corresponding graphs must then include two vertices in the list (\ref{eq:tlv}).

Operators of dimension 6 are much more plentiful: for a single family, assuming 
lepton and baryon number conservation and a single scalar doublet, they number 
82~\cite{bw}. The list of \tlgo s is shorter, though still numerous (45 operators);
the generic forms of these operators are
\beq
\phi^6 , \qquad D^2 \phi^4 , \qquad 		
\psi^2 \phi^3, \qquad D \psi^2 \phi^2, \qquad 	
(\bar\psi \psi)^2, \qquad (\bar\psi\gamma^\mu\psi)^2	
\label{eq:tlgo}
\eeq
where $\phi,~\psi$ and $D$ denote scalars, fermions and covariant
derivatives respectively. The detailed list of operators
is presented in Ref. \cite{patterns},\footnote{It must be emphasized that 
(\ref{eq:tlgo}) represents operators that {\em might} be generated by an 
underlying gauge theory taking into account only the 
most general constraints imposed by gauge invariance. Specific
models might contain additional restrictions that prevent the appearance of
one or more \tlgo s. In contrast, operators {\em not} in this list cannot be
generated at tree-level by {\em any} gauge theory (triple-vector-boson
operators fall in this category).} where the graphs responsible for 
their generation is also diplayed. It is then a simple matter to
verify that the absence of the vertices (\ref{eq:tlv}) excludes 
the operators (\ref{eq:tlgo}).

In this case, however, there is a simplification: none of the dimension 6
effective operators continaing fermions and vector-bosons (but no scalars)
can be generated at tree level~\cite{patterns}~\footnote{The reason is that
due to gauge invariance an operator involving only vectors and fermions 
such as $ \bar \ell \sigma^I \gamma^\mu D^\nu \ell W^I_{\mu\nu} $ (where 
$W^I_{\mu\nu}$ denotes the $\su2$ field strength), must
contain a term with only three fields, and no tree-level graph with 3 external
legs and $\ge1$ internal lines can be contructed.} and because of this the
vertex of type $ \Psi \psi A $ in (\ref{eq:tlv}) is redundant when considering
oeprators of dimension $\le6$.

\section{Elimination of \tlgo s}
\label{sect:eliminate}

The minimal statement leading to the absence of \tlgo s is the one made above,
namely, that all vertices in (\ref{eq:tlv}) be absent. This can be achieved in a
natural way by assuming that the underlying theory has certain symmetry 
properties.

The simplest case correpsonds to that where the heavy fields are all non-singlets
under a certain symmetry while all light fields are invairant. This is 
realized in the MSSM~\cite{mssm} provided we label both scalar doublets as light,
(the ``new'' symmetry is then essentially R-parity~\cite{r.par}).
In this case  $ \Lambda $ is set by the scale of the soft-breaking terms and
by the dimensional coefficient in the superpotential.

Another example is provided by the so called ``universal'' higher-dimensional
theories~\cite{ued}. In these models space time is assumed to have $4+\delta $
dimensions such that its ground state consists of the topological
product of our non-compact Euclidean 4-dimensional space-time and a compact
Riemannian manifold $\rcal $ of dimension $ \delta $. All heavy modes are associated
with a non-zero momentum in the directions corresponding to $ \rcal $, while all
light (\sm) particles have zero momenta along these directions.
As a result there are no vertices containing a single heavy mode.

\subsection{Constraints on the underlying local symmetry}

It is also possible to arrange for the local symmetry of the full theory to
eliminate the vertices (\ref{eq:tlv}), but the models are more convoluted.
I will therefore present only a simplified discussion not intended to generate 
realistic theory. Specifically, I will consider a generic renormalizable gauge 
with scalar fields $ s $, left and right-handed femrions $ \chi_{L,R} $
and gauge bosons $V$, whose local symmetry  group $ \gcal $ is broken
to $ \hcal $ at scale $ \Lambda $. I will make no attempt to insure that the
light theory correpsonds to the \sm, not will I include the interactions that
generate masses for the light excitations~\footnote{Light masses are presumably generated
by introducing other scalar field(s). As usual there will be a hierarchy
problem~\cite{hierarchy} when attempting to maintain the gap between $ \Lambda $ 
and the low energy scale.} 

I will assume that {\em all}  scalar $ \gcal$-multiplets receive a \vev\  
$\langle s \rangle = O( \Lambda) $, that all the heavy fermions and
vector-boson masses are generated in this manner, and that all the resulting 
physical scalars are heavy. This last condition would not be realized in cases where the scalar 
potential for the $ s $ has a symmetry group larger than $ \gcal $; this 
type of models are explicitly excluded.

These constraints, together with the fact that the unbroken generators 
annihilate $ \langle s \rangle$ and that the generators of $ \hcal $ 
close into an algebra, eliminate all the vertices in ({\ref{eq:tlv})
except thos of the form  $\Psi \psi A$ and $ X \psi \psi $.
In particular, vertices of the form $ \Phi \psi \psi $ are excluded by the
condition that all fermions receive a mass through \ssb\ and that all scalar
multiplets receive a \vev\ of order $ \Lambda $. 

The remaining  $\Psi \psi A$ and $ X \psi \psi $ vertices can be eliminated only
by appropriate choice of the representations for the scalar and fermion fields.
Note that if one is interested only in suppressing the leading heavy physics efects
as generated by operators of dimension $\le6$ only vertices of type $ \psi\psi X $ need
be eliminated, see section~\ref{sect:dimvvi}. I will consider several possiblities 
when $ \gcal = \su N $.
\bit
\item If $ \chi_L $ transform according to the fundamental representation
and $ \chi_R$ is a singlet, or if $ \chi_L $ carries the adjoint and $ \chi_R$
the fundamental, and if in either case $s$ carries the fundamental representation, 
then $ \hcal = \su{N-1} $, and vertices of the form
$\Psi \psi A$ and $ X \psi \psi $ are asbent given this pattern
of symmetry breaking (most easily seen by choosing a gauge where 
$ \langle s \rangle = ( 0, \ldots, 0,v) $).
\item This is also trivially the case in a vector-like theory where
$ \chi_{L,R} $ transform accoding to the fundamental reprsetnation and $s$
in the adjoint, for in this case there are no light scalars.
\item If both $ \chi_L$ and $ s $ carry the adjoint representation and $ \chi_R $ 
is a singlet, then $ \hcal = \su{[N/2]} \otimes \su{N-[N/2]} \otimes \ui $\cite{li}, 
and though vertices of the type $\Psi \psi A$ are excluded, those of the form $ X \psi \psi $
will be present.
\eit
Theories where vertices of type $ \psi \psi X $ are not eliminated will
exhibit realatively large heavy physics effects at low energies in the
form of chirality-preserving four-fermion interactions.

\begin{table}
\begin{tabular}{|c|c|c|}
\hline
heavy particle & weak isospin & $|$hypercharge$|$        \cr \hline\hline
vector, scalar       & $0,~1$     & $n/3,~0\le n \le 5 $ \cr
scalar               & $1/2,~3/2$ & $1/2,~3/2 $          \cr
left-handed fermion  & $0,~1$     & $n/3,~0\le n \le 3 $ \cr
right-handed fermion & $1/2$      & $1/6,~1/2$           \cr \hline
\end{tabular}
\caption{The $\su2_L\times\ui_Y$ transformation properties of the heavy particles
that would allow the corresponding vertices in (\ref{eq:tlv}) to occur.}
\label{tab:su2ui}
\end{table}

\subsection{Constraints form $\su2_L\times \ui_Y$}

As a final example
I will consider the conditions under which the \sm\ gauge symmetry
eliminates all the vertices in (\ref{eq:tlv}). This can be implemented
by requiring that the none of the heavy fields in each term in (\ref{eq:tlv})
carries the same $\su3\times\su2\times\ui$ representation as the factor 
containing light fields. For example, when
considering the vertex $ \psi \psi X $, the product $ \psi \psi $ is 
either a fermion triplet of hypercharege $0$ or $\pm1/3$, or an
$\su2$ singlet of hypercharge $ \pm n/3, ~0\le n\le 5 $ (assuming the
vertex odes not violate B-L and the presence of right-handed neutrinos).
Then $ \psi \psi X $ is excluded if no $X$ carries any of these
representations.
Table \ref{tab:su2ui} all the representations excluded in this way

Note in particular that if the underlying theory is a spontaneously
broken gauge theory and if absence of the vertices
$ \psi \psi X $ and $ \phi \phi X$ is a consequence of the \sm\ gauge
symmetry, then the underlying theory cannot contain a heavy vector-boson
which is a \sm\ singlet. But the absence of such heavy
vector-boson singlets implies that none of the group generators broken
at scale $ \Lambda $ commute with those of $\su3\times\su2_L\times \ui$, and this 
implies that the \sm\ $\su3_c \times \su2_L \times \ui_Y $ is the gauge group
of the {\em full} theory.
%
%
%

\section{Numerical considerations and conclusions}
\label{sect:end}

The above arguments indicate that for weakly-coupled and decoupling
theories there are (at least) two ways of hiding the heavy physics
effects from current data. The first simply assumes that 
the scale of new physics is very large, thereby suppressing 
all heavy particle effects. The second assumes the absence 
of the vertices in (\ref{eq:tlv}) (protected by a symmetry)
so as to strongly suppress all virtual effects,
and a moderately high value for $ \Lambda $ in order to explain 
the absence of direct heavy particle production.

These two approaches are not necessarily mutually exclusive. 
For example, we can assume that all effective operators that involve
lepton-number violation are generated at scale
$ \Lambda_{\not L}$, while all 
operators that respect all discrete symmetries of the
\sm\ are generated by physics with typical
scale $ \Lambda \ll \Lambda_{\not L}$. Then lepton-number violating effects
can be suppressed by having $ \Lambda_{\not L}^2 G_F \gg 1$,
while the effects of operators generated at scale $ \Lambda $
can be suppressed by insuring the absence of the vertices in
(\ref{eq:tlv}), where the quantum numbers involved in
each such vertex respects all the discrete symmetries of the \sm\
(this assumes that appropriate transformation properties can be 
assigned to the heavy particles).

Assuming the absence of tree-level effects one might reasonably 
ask how would existing limits be modified for models of the 
type considered in this paper. To estimate this note that 
the surviving effects occur via loops and 
appear with an additional factor of 
$ \sim 1/(4\pi)^2 $, so that a  limit $ \Lambda > \mexp $,
obtained when assuming the presence of \tlgo s,
is replaced by $ \Lambda > \mexp/(4\pi)$: the existing 
limit is decreased by an order of magnitude. 
For processes that are not suppressed within the \sm\
the contributions from the heavy physics are suppressed
by a factor $ \sim (m_W/\Lambda)^2 $ with respect to the
\sm\ {\em radiative} corrections. Sensitivity to such effects
requires a precision below $0.01\%$.

As a specific example consider the limits on ``compositeness'' obtained
using operators
of the form $ (\qwe/\Lambda^2) \left( \bar \psi_1 \gamma_\mu \psi_2 \right)
\left( \bar \psi_3 \gamma_\mu \psi_4 \right)$, involving light left-handed 
fermions~\cite{pdg}. For example, for left-handed electrons,
the exisiting limit is $ \Lambda \gesim 10 \tev $
obtained using $ \qwe = 2\pi^2 $; in the absence of \tlgo s, using
loop-generated operators for which $ \qwe \sim 1/(4\pi^2) $, this is relaxed to
$ \Lambda > 360\gev$. 
Similarly, the limit on $\Lambda $ presented in~\cite{pdg} 
for all 4-fermion (current-current) operators are decreased by a factor $ \sim 30$.

These results imply that there is a class of heavy-physics models for which all leading
virtual effects can be naturally suppressed, and this leads to a considerable relaxation 
of the existing limits on the scale of new physics (by an order or magnitude or more).
The simplest way of realizing this situation is to assume that all heavy excitations
transform non-trivially under a certain (unknown) symmetry under which all \sm\
particles are singlets; which also implies that that near future accelerators will 
only pair-produce these heavy particles. Should this class of models describe the physics
underlying the \sm, experiments will have at most weak indications of the presence
and properties of any heavy excitations before they are directly produced.

\appendix

\section{Absence of vertices of type $XA \phi \phi $}

It is worth noting that the vertex $XA \phi \phi $, 
is not included
since it follows form the absence of vertex $ X \phi \phi $.
To see this collect all the scalars
in a large vector $ \chi $ such that $ \chi^\dagger = ( \phi^\dagger,
\Phi^\dagger ) $ (where $ \Phi $ denotes the heavy scalar fields).
Denote now the group generators for the (in general reducible)
representation carried by $ \chi $ by $ \{ T_a \}$, of which $ \{ T_I \}$ generate
the \sm\ gauge group while $ \{ T_r \}$  are the remaining ones.
I can then choose
\beq
T_I = \pmatrix{\tau_I & 0 \cr 0 & t_I} \qquad
T_r = \pmatrix{U_r & N_r \cr N_r^\dagger & M_r }
\eeq
since the \sm\ scalars transform irreducibly under the \sm\ gauge group.\\
Now, the scalar kinetic energy is
\beq 
| \partial \chi + i T_a V^a \chi |^2,
\eeq
where $V$ denotes all the gauge fields, $X$ or $A$, and where 
the gauge couplings have been absorbed into the gauge fields.
Then the $X\phi\phi$ and $XA\phi\phi$ vertices are explicitly given by
\beq
X^r_\mu \cdot \phi^\dagger {\stackrel\leftrightarrow \partial^\mu} U_r \phi
\qquad
\left(X^r \cdot A^I \right) \phi^\dagger \{ U_r , \tau_I \}\phi
\eeq
The first will be absent provided the matrix  $ U_r $ vanishes, 
but this then implies that the vertex $XA \phi \phi $  is also absent.


\begin{thebibliography}{99} 
\bibitem{limits} 
J.~Drees,
Int.\ J.\ Mod.\ Phys.\ A {\bf 17}, 3259 (2002)
[arXiv:hep-ex/0110077].
%
F.~Parodi  [LEP Collaborations],
AIP Conf.\ Proc.\  {\bf 618} (2002) 32.
%
G.~Altarelli,
arXiv:hep-ph/0306055.
%
\bibitem{limits.models}
See, for example, 
P.~Langacker, M.~x.~Luo and A.~K.~Mann,
Rev.\ Mod.\ Phys.\  {\bf 64}, 87 (1992).
%
\bibitem{limits.leff} 
See, for example, 
J.~Ellison and J.~Wudka,
Ann. Rev. Nuc. Part. Sci. {\bf 48}, 33 (1998)
[arXiv:hep-ph/9804322].
%
S.~Alam, S.~Dawson and R.~Szalapski,
Phys.\ Rev.\ D {\bf 57}, 1577 (1998)
[arXiv:hep-ph/9706542].
%
\bibitem{pdg}
K.~Hagiwara {\it et al.}  [Particle Data Group Collaboration],
Phys.\ Rev.\ D {\bf 66}, 010001 (2002).
%
\bibitem{patterns}
C.~Arzt, M.~B.~Einhorn and J.~Wudka,
Nucl.\ Phys.\ B {\bf 433}, 41 (1995)
[arXiv:hep-ph/9405214].
%
\bibitem{decoupling}
T.~Appelquist and J.~Carazzone,
Phys.\ Rev.\ D {\bf 11}, 2856 (1975).
%
J.~C.~Collins, F.~Wilczek and A.~Zee,
Phys.\ Rev.\ D {\bf 18}, 242 (1978).
%
\bibitem{non.dec}
For a recent review see:
 A.~Dobado and M.~T.~Urdiales,
Z.\ Phys.\ C {\bf 71}, 659 (1996)
[arXiv:hep-ph/9502255].

\bibitem{leff}
S.~Weinberg,
PhysicaA {\bf 96}, 327 (1979);
%
arXiv:hep-th/9702027.
H.~Georgi,
Nucl.\ Phys.\ B {\bf 361}, 339 (1991);
%
Nucl.\ Phys.\ B {\bf 363}, 301 (1991).
%
J.~Wudka,
Int.\ J.\ Mod.\ Phys.\ A {\bf 9}, 2301 (1994)
[arXiv:hep-ph/9406205].
%
\bibitem{weinberg}
S.~Weinberg,
Phys.\ Rev.\ Lett.\  {\bf 43}, 1566 (1979).
%
\bibitem{bw}
W.~Buchmuller and D.~Wyler,
Nucl.\ Phys.\ B {\bf 268}, 621 (1986).
%
W.~Buchmuller, B.~Lampe and N.~Vlachos,
Phys.\ Lett.\ B {\bf 197}, 379 (1987).
%
\bibitem{nat}
G.~'t Hooft,
PRINT-80-0083 (UTRECHT)
{\it Lecture given at Cargese Summer Inst., Cargese, France, Aug 26 - Sep 8, 1979}
%
\bibitem{mssm}
H.~P.~Nilles,
Phys.\ Rept.\  {\bf 110}, 1 (1984).
%
H.~E.~Haber and G.~L.~Kane,
Phys.\ Rept.\  {\bf 117}, 75 (1985).
%
R.~Barbieri,
Riv.\ Nuovo Cim.\  {\bf 11N4}, 1 (1988).
%
J.~F.~Gunion and H.~E.~Haber,
Nucl.\ Phys.\ B {\bf 272}, 1 (1986)
[Erratum-ibid.\ B {\bf 402}, 567 (1993)];
Nucl.\ Phys.\ B {\bf 278}, 449 (1986);
Nucl.\ Phys.\ B {\bf 307}, 445 (1988)
[Erratum-ibid.\ B {\bf 402}, 569 (1993)].
%
J.~Rosiek,
Phys.\ Rev.\ D {\bf 41}, 3464 (1990).
%
\bibitem{r.par}
P. Fayet and J. Iliopoulos, 
Phys. Lett. {\bf 51}, 461 (1974).
%
G. Ferrar and P. Fayet,
Phys. Lett. {\bf 76B}, 575 (1978),
{\it ibid.} {\bf 79B}, 442 (1978).
%
\bibitem{ued}
T.~Appelquist, H.~C.~Cheng and B.~A.~Dobrescu,
Phys.\ Rev.\ D {\bf 64}, 035002 (2001)
[arXiv:hep-ph/0012100].
%
\bibitem{hierarchy}
See, for example,
E.~Farhi and L.~Susskind, 
Phys.\ Rept.\ {\bf 74}, 277 (1981). 
%
P.~Langacker,
Phys.\ Rept.\  {\bf 72}, 185 (1981).
%
\bibitem{li} 
L.~F. Li
Phys.\ Rev.\ D {\bf 9}, 1723 (1973)
%
\end{thebibliography}
\end{document}